\documentclass[prl,reprint,amsmath,amssymb,showpacs,superscriptaddress,floatfix,longbibliography]{revtex4-1}
\usepackage[breaklinks=true,colorlinks,citecolor=blue,linkcolor=blue,urlcolor=blue]{hyperref}
\usepackage{graphicx,epsfig}
\usepackage{amsmath}
\usepackage{xcolor}
\usepackage{color}
\renewcommand{\BibitemShut}[1]{}

\def\be{\begin{equation}}
\def\ee{\end{equation}}

\def\bea{\begin{eqnarray}}
\def\eea{\end{eqnarray}}

\begin{document}

\title{Thermal and gravitational chiral anomaly induced magneto-transport in Weyl semimetals}
\author{Kamal Das}
\email{kamaldas@iitk.ac.in}
\affiliation{Dept. of Physics, Indian Institute of Technology Kanpur, Kanpur 208016, India}
\author{Amit Agarwal}
\email{amitag@iitk.ac.in}
\affiliation{Dept. of Physics, Indian Institute of Technology Kanpur, Kanpur 208016, India}

\begin{abstract}
Quantum anomalies in Weyl semimetal (for either ${\bf E}\cdot{\bf B} \neq 0$ or ${\bf \nabla}T\cdot{\bf B} \neq 0$) leads to chiral charge and energy pumping between the opposite chirality nodes.  
This results in chiral charge and energy imbalance between the Weyl nodes which manifests in several intriguing magneto-transport phenomena. Here, we investigate the role of electrical-, thermal-, and gravitational chiral anomaly on magneto-transport in Weyl semimetals. We predict the planar Ettinghausen and Righi-Leduc effect to be a distinct signature of these quantum anomalies.
We also demonstrate a significant enhancement in the thermo-electric conductivity, Seebeck effect, Nernst effect and thermal conductivity with increasing temperature. Interestingly, this anomaly induced transport violates the Wiedemann-Franz law and Mott relation. 
\end{abstract}
\maketitle

Massless chiral fluids in presence of a magnetic field exhibit quantum anomalies, which manifest as the non-conservation of chiral charge and energy densities \cite{Nielsen83,Landsteiner11,Son13,Lucas16,Gooth17,PhysRevLett.120.206601}. Weyl semimetals (WSM) hosting a pair of Weyl nodes of opposite chirality, which also act as Berry curvature monopoles \cite{Wan11, Armitage18, Xu613, Huang15a, Soluyanov15, Hasan17}, 
offer an ideal platform to explore these. The non-conservation of chiral charge in presence of electric field (${\bf E}\cdot {\bf B} \neq 0$) is known as electrical chiral anomaly (ECA) \cite{Nielsen83,Son12, Son13}. This leads to very interesting magneto-electric transport phenomena~\cite{Son13,Burkov12,Yip15,Xiong15,Huang15b,Lucas16,Li16a,Li16b,
Burkov17,Nandy17} in WSM. 
The non-conservation of chiral energy in presence of temperature gradient (${\bf \nabla}T \cdot {\bf B} \neq 0$) is a manifestation of the {gravitational chiral anomaly} (GCA) \cite{Landsteiner11,Lucas16,Gooth17}. This leads to interesting signatures in magneto-thermal transport in WSM \cite{Lundgren14, Kim14,Hirschberger16, Jia16,Spivak16, Sharma16,Stockert17, Gooth17,
 Zyuzin_V_A17,Sharma18, Nandy19}. Here, we demonstrate another kind of chiral anomaly associated with chiral charge pumping in presence of ${\bf \nabla}T \cdot {\bf B} \neq 0$, the thermal chiral anomaly (TCA). 

\begin{figure}[t]
\includegraphics[width=0.95\linewidth]{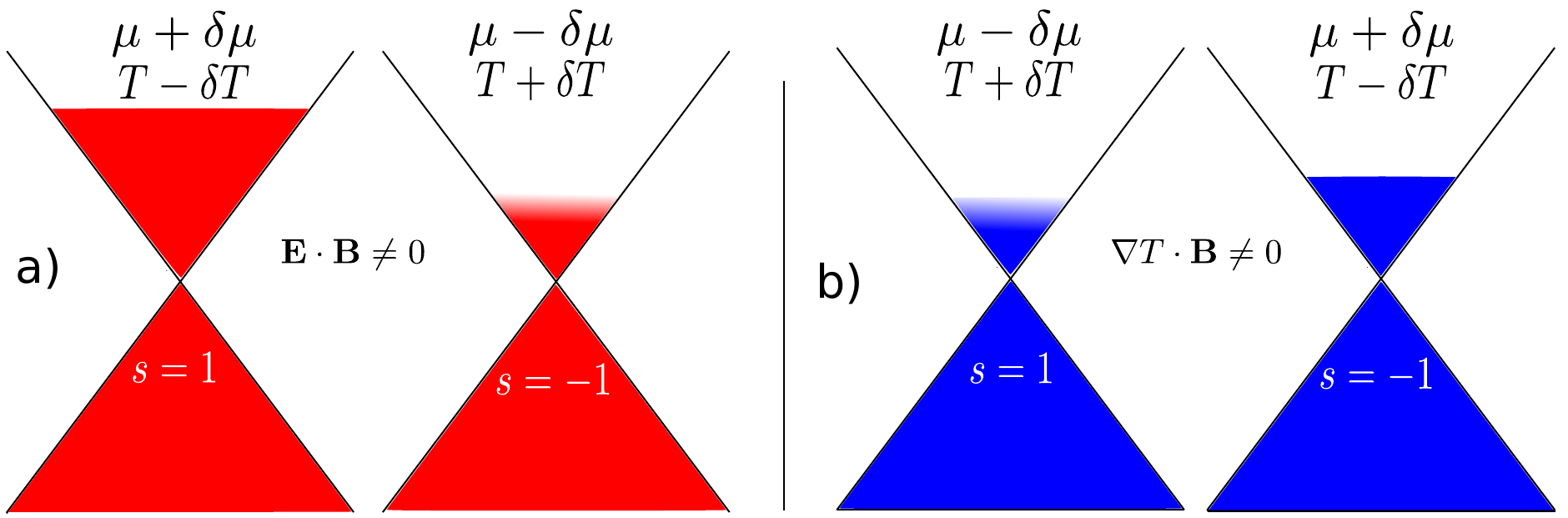}
\caption{Schematic of the chiral chemical potential ($\delta \mu$) and chiral temperature ($\delta T$) imbalance in WSM for a) ${\bf E}\cdot{\bf B} \neq 0$ and 
(b) ${\bf \nabla}T\cdot{\bf B} \neq 0$. Both of these lead to quantum chiral anomalies, which pump chiral charge and energy from one Weyl node to the other. 
\label{fig_1}}
\end{figure}

The origin of these quantum chiral anomalies can be traced to the magnetic field induced {\it equilibrium} chiral charge and energy current in WSM (${\bf j}^s_{e,{\rm eq}}$ and ${\bf j}^s_{{\cal E},{\rm eq}}$, respectively, with $s = \pm1 $ being the chirality). 
These are 
%
%
\bea \label{chrg}
{\bf j}^s_{e, {\rm eq}} &=& - e \left(\mu{\cal C}_0^s  {\bf B} + T {\cal C}_1^s  {\bf B}\right)~,\\\label{heat}
{\bf j}^s_{{\cal E},{\rm eq}} &=& \mu^2 \dfrac{ {\cal C}_0^s}{2} {\bf B} + \mu T {\cal C}_1^s{\bf B} + T^2 {\cal C}_2^s {\bf B}~.
\eea
Here, ${\cal C}_i^s$ (defined later) are the coefficients of different quantum chiral anomalies. 
Equation $\eqref{chrg}$ generalizes the chiral magnetic effect in WSM to include finite temperature. In Eq.~\eqref{chrg}, ${\cal C}_0^s$ is the ECA coefficient, 
while ${\cal C}_1^s$ determines the charge pumping in presence of a finite $T$. It has not been explored earlier and we will refer to it as the coefficient of TCA. In Eq.~\eqref{heat} the first two terms simply denote the energy carried by chiral charge current, while the third term $\propto {\cal C}_2^s$ captures the thermal component and it is known to be analogous to the GCA \cite{Landsteiner11,Lucas16,Gooth17}. See 
Sec.~S1 of the supplementary materials (SM)  
\footnote{\href{https://www.dropbox.com/s/fdw4xhmdlbq4mia/THE_supp_v2.pdf?dl=0}{Supplementary material} detailing the following: 1) The chiral charge and energy current in equilibrium and coefficients of the quantum anomalies, 
2) Local equilibrium approximation, 3) Chiral chemical potential and temperature imbalance, 4) Anomaly induced transport coefficients, 5) Berry curvature and 
magnetic moment, and 6) Temperature dependence of transport coefficients, and 7) Violation of Wiedemann Franz law and Mott relations 8) Sign reversal of $\alpha$.}
for more details. 
In a {\it non-equilibrium} scenario, this chiral charge pumping is stabilized by inter-node scattering and results in 
chiral charge (different $\mu^s$) and energy imbalance (different $T^s$) in the two Weyl nodes (see Fig.~\ref{fig_1}). This charge and energy imbalance in the Weyl nodes, gives rise to 
several interesting effects in magneto-transport experiments. 

In this letter, we present a unified framework for these three chiral quantum anomalies in the Boltzmann transport formalism. We explicitly calculate all the magneto-transport coefficients, and predict their magnetic field dependence, angular dependence (between ${\bf B}$ and ${\bf E}$ or ${\nabla T}$), and temperature scaling. In addition to the planar Hall and planar Nernst effect, we predict planar Ettinghausen and Righi-Leduc effect to be a manifestation of these anomalies. 
Remarkably, we find significant enhancement in the magnetic field induced thermo-electric conductivity, Seebeck and Nernst effects and thermal conductivity 
with increasing temperature. We also demonstrate that the quantum chiral anomaly induced transport coefficients violate the Wiedemann-Franz law, as well as the Mott relation in WSM. 

The dynamics of charge carriers in a Weyl cone of a given chirality is described by the following equations for the carriers position ($\bf r$) and momentum ($\bf k$)~\cite{Sundaram99, Xiao10}, 
\begin{eqnarray}\label{eom_r}
\dot{\bf r}^s & = &\Delta^s\left[\tilde{\bf v}^s +\frac{e}{\hbar}{\bf E}\times {\bf \Omega}^s +\frac{e}{\hbar}(\tilde{\bf v}^s \cdot {\bf \Omega}^s){\bf B}\right]~,
\\\label{eom_k}
\hbar\dot{\bf k }^s &=& \Delta^s \left[-e{\bf E} - e\tilde{{\bf v}}^s\times {\bf B}-\frac{e^{2}}{\hbar}({\bf E}\cdot{\bf B}){\bf  \Omega}^s\right]~.
\end{eqnarray}
Here, `$-e$' is the electronic charge, ${\bf  \Omega}^s$ is the Berry curvature, and $\Delta^s \equiv 1/(1 + \frac{e}{\hbar}{\bf \Omega}^s \cdot {\bf B})$ is the phase-space factor~\cite{Xiao05}. 
In presence of magnetic field, the energy dispersion of the carriers is modified to include the orbital magnetization, $\tilde \epsilon^s = \epsilon^s - {\bf m}^s \cdot {\bf B}$. The resultant band velocity is, $\tilde{\bf v}^s= \frac{1}{\hbar} \frac{\partial \tilde \epsilon^s}{\partial {\bf k}}$ \cite{Xiao06}.
In Eq.~\eqref{eom_r},  
the 
third term $(\tilde{\bf v}^s\cdot {\bf \Omega}^s){\bf B}$ is the chiral magnetic velocity \cite{Son12, Burkov12} which gives rise to the equilibrium chiral charge and energy currents in Eqs.~\eqref{chrg}-\eqref{heat}.
It is the foundation for all three quantum chiral anomalies discussed in this letter.
%


In presence of external perturbations, the non-equilibrium distribution function ($g_{{\bf r},{\bf k}}$) for each node is given by~\cite{Ashcroft76}, 
\be\label{bte_1}
\dfrac{\partial g_{{\bf r},{\bf k}}^s}{\partial t} + \dot{\bf r}^s \cdot {\bf \nabla}_{\bf r } ~g_{{\bf r},{\bf k}}^s +\dot{\bf k }^s \cdot {\bf \nabla}_{\bf k}~g_{{\bf r},{\bf k}}^s = I_{\rm coll} \{g_{{\bf r},{\bf k}}^s\}~.
\ee
To reach a steady state in presence of chiral charge and energy pumping between the two Weyl nodes, the collision integral ($I_{\rm coll}$) should include both the intra-node ($\tau_0$) as well as the inter-node ($\tau_v$) scattering timescales. Furthermore, owing to the chiral charge and energy pumping, each of the Weyl nodes   
is assumed to acquire a {\it local equilibrium} (LE) chemical potential $\mu^s \equiv \mu + \delta \mu^s$ \cite{Yip15,Das19a} and temperature $T^s = T + \delta T^s$. Within this approximation the steady state collision integral is given by 
\be \label{I_coll_3}
I_{\rm coll}^s= -\dfrac{g^s_{{\bf r},{\bf k}}-f\left(\tilde \epsilon^s,\mu^s,T^s\right)}{\tau_0} - \dfrac{g^s_{{\bf r},{\bf k}} - f(\tilde \epsilon^s,\mu^{\bar s},T^{\bar s})}{\tau_{v} }~.
\ee 
Here, $f\left(\tilde \epsilon^s,\mu^s,T^s\right)$ is the Fermi function with energy $\tilde\epsilon^s$, chemical potential $\mu^s$ and temperature $T^s$.
The first term in Eq.~\eqref{I_coll_3} reflects the relaxation of $g^s_{{\bf r},{\bf k}}$ to the LE of the same node via intra-node scattering while the second term specifies its relaxation to the LE of the other node by inter-node scattering. For simplicity we assume the systems to have a small Fermi surface so that the energy or momentum dependence of 
$\tau_0$ and $\tau_v$ can be ignored. 
Additionally, we work in the `chiral limit', $\tau_v \gg \tau_0$, where the transport is dominated by the inter-node scattering. 


Substituting this collision integral in Eq.~\eqref{bte_1}, and integrating over all momentum modes, we obtain the following equation 
for the particle number (${\cal N}^s$) dynamics, 
%
\be\label{cont}
\dfrac{\partial {\cal N}^s}{\partial t} 
+{\cal C}_{0}^s~e{\bf E} \cdot{\bf B} + {\cal C}_{1}^s{\nabla T} \cdot{\bf B} = \dfrac{{\cal N}^s -{\cal N}^{\bar s}}{\tau_v}.
\ee
%
This generalizes the semiclassical ECA ($\propto {\cal C}_{0}^s$)~\cite{Son13} to include the temperature gradient induced chiral charge pumping, or TCA ($\propto {\cal C}_{1}^s$). Similarly we 
calculate the energy dynamics to be,  %
\be  \label{CA_enrgy_eqn}
\dfrac{\partial {\cal E}^s}{\partial t} + (\mu {\cal C}^s_0 +T {\cal C}^s_1)~e{\bf E} \cdot{\bf B} + (\mu  {\cal C}_1^s + 2 T{\cal C}_2^s)~{\nabla T} \cdot{\bf B} = \dfrac{{\cal E}^s -{\cal E}^{\bar s}}{\tau_v}.
\ee
Here, the $\mu {\cal C}^s_0$ and the $\mu  {\cal C}_1^s$ terms represent the energy carried by the chiral charge transfer. The 
$2 T{\cal C}_2^s$ term highlights the energy pumped by the $\nabla T\cdot {\bf B}\neq 0$ term and it 
is analogous to GCA. 
%
%
%

\begin{figure}[t]
\includegraphics[width=.95\linewidth]{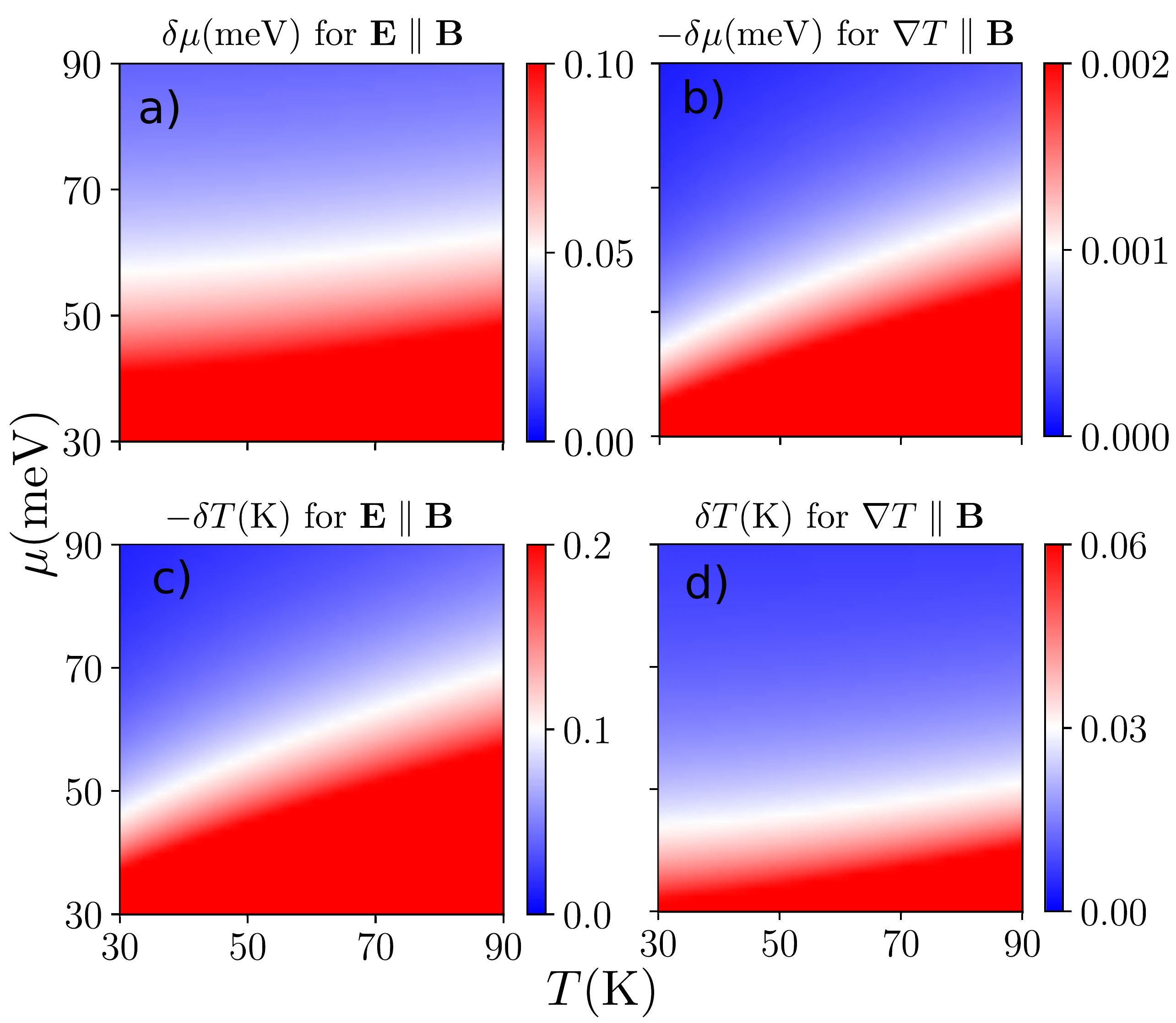}
\caption{The anomaly induced chiral chemical potential $\delta \mu^s$ and chiral temperature $\delta T^s$ for the $s =1$ node, in the $\mu-T$ plane. The $\delta \mu^s$ induced by (a)  ${\bf E} \parallel {\bf B}$ and (b) ${\bf \nabla}T \parallel {\bf B}$ have opposite signs. 
The $\delta T^s$ induced by (c)  ${\bf E} \parallel {\bf B}$ and (d) ${\bf \nabla}T \parallel {\bf B}$, also have opposite signs. 
In all cases, the impact of the ${\bf E} \parallel {\bf B}$ term is larger. Furthermore, while the ${\bf E}\cdot {\bf B} \neq 0$ term increases the chiral $\delta \mu$ of the positive chirality node, it reduces its chiral temperature $\delta T$. Here we have chosen $v_F = 2 \times 10^5$ m/s, $\tau_v = 10^{-9}$ s, $B = 6$ T, sample length $l = 50~\mu$m, $|{\bf E}| = 1$ mV/$l$, and $|\nabla T| = 350$ mK/$l$ \cite{Gooth17}. 
\label{fig_2}}
\end{figure}

Working in the linear response regime in ${\bf E}$ and $\nabla T$, we find that the imbalance of the chiral carriers and temperatures are small, {\it i.e.}, $\delta \mu^s < \mu$, and $\delta T^s< T$. Solving for $\delta \mu^s$ and $\delta T^s$ (see Sec. S2 and S3 in SM \cite{Note1} for more details) in this regime, we obtain 
%
%
\be \label{chrl_qnts}
\begin{pmatrix}
\delta \mu^s \\
 \delta T^s/T
\end{pmatrix}
= -\dfrac{\tau_v}{2}
\begin{pmatrix}
{\cal D}_0^s & {\cal D}_1^s\\
{\cal D}_1^s & {\cal D}_2^s
\end{pmatrix}^{-1}
\begin{pmatrix}
{\bf \Lambda}_0^s & {\bf \Lambda}_1^s\\
{\bf \Lambda}_1^s & {\bf \Lambda}_2^s
\end{pmatrix}
\cdot 
\begin{pmatrix}
e{\bf E} \\
{\nabla T}/T
\end{pmatrix}~.
\ee
Here, we have defined the generalized {\it energy densities} and the generalized {\it energy velocities} at the Fermi energy 
\be \label{DnVn}
\begin{pmatrix}
{\cal D}_n^s \\
{{\bf \Lambda}}_n^s\end{pmatrix} = \int \frac{d{\bf k}}{(2\pi)^3} (\tilde \epsilon^s - \mu)^n \left(-\partial_{\tilde \epsilon^s} f\right) 
\times \begin{pmatrix}
1+e{\bf \Omega}^s{\bf B}/\hbar\\
\tilde{\bf v}^s  + \frac{e}{\hbar}(\tilde{\bf v}^s \cdot {\bf \Omega}^s){\bf B}
\end{pmatrix}, 
\ee
with $n = \{0,1,2\}$.  
Here, ${\cal D}_0^s$ defines the finite temperature density of states in presence of Berry curvature. ${\bf \Lambda}_0^s$ and ${\bf \Lambda}_1^s$ are the total chiral magnetic velocity and the total chiral energy velocity at the Fermi level, respectively. It is evident from Eq.~\eqref{chrl_qnts} that both the electric field and the temperature gradient contribute to generate the chiral chemical potential and chiral temperature imbalance in the WSM. It turns out that in ${\bf \Lambda}_n^s$, only the contribution of the {\it chiral magnetic velocity} survives and we have ${\bf \Lambda}_n^s  = \Lambda_n^s {\bf B}$, along with $\Lambda_n^s \propto s$. 

The anomaly coefficients are explicitly given by, 
\be
\left\{{\cal C}_0^s, {\cal C}_1^s, {\cal C}_2^s \right\} = \left\{ \Lambda_0^s, \frac{{\Lambda}_1^s}{T}, \frac{{\Lambda}_2^s}{2T^2} \right\}~. 
\ee
Clearly, in this semiclassical formalism, all these three anomalies arise from the presence of a finite Berry curvature and chiral magnetic velocity. However, while the chiral anomaly 
($\propto {\cal C}_0^s$) is well explored in WSM, the other two anomalies of a similar origin are relatively less explored [\onlinecite{Lucas16, Gooth17}]. 
Developing a common theoretical framework to explore all of them is one of the main highlights of this letter. 


The anomaly induced contribution to the charge (${\bf j}_e =\sum_s {\bf j}_e^s$) and energy (${\bf j}_Q =\sum_s {\bf j}_Q^s$) current can now calculated to be 
\small 
\be \label{currents}
\begin{pmatrix}
{\bf j}_e^s \\
{\bf j}_Q^s
\end{pmatrix}
= 
\dfrac{{\bf B}\tau_v}{2}
\begin{pmatrix}
e{\Lambda}_0^s & e{\Lambda}_1^s\\
-{\Lambda}_1^s & -{\Lambda}_2^s
\end{pmatrix}
{{\cal D}_s}^{-1}
\begin{pmatrix}
{\Lambda}_0^s & {\Lambda}_1^s\\
{\Lambda}_1^s & {\Lambda}_2^s
\end{pmatrix}
\begin{pmatrix}
e {\bf B \cdot E} \\
{{\bf B \cdot \nabla}T/T}
\end{pmatrix},
\ee
\normalsize
where \small 
${\cal D}_s = \begin{pmatrix}
{\cal D}_0^s & {\cal D}_1^s\\
{\cal D}_1^s & {\cal D}_2^s
\end{pmatrix}$. \normalsize
Now, as $\Lambda_n^s \propto C_n^s$, Eq.~\eqref{currents} implies that the charge and energy current in WSM are associated with ECA, TCA, GCA, or mixed anomalies (a combination of two of these). Thus Eq.~\eqref{currents} captures the essence of the quantum anomalies induced magneto-transport in WSM, and is one of the main results of this work. See Sec. S4 in the SM \cite{Note1} for more details, and expanded version of Eq.~\eqref{currents}.
%

The transport coefficients can be obtained by comparing Eq.~\eqref{currents} to the phenomenological linear response,  
$j_{e,i} = \sum_j [\sigma_{ij}~E_j - \alpha_{ij}~\nabla_j T$] and $ j_{Q,i} = \sum_j [{\bar \alpha}_{ij}~E_j - {\bar \kappa}_{ij}~\nabla_j T ]$. 
Here, $\sigma$, $\alpha$, $\bar\alpha$ and $\bar \kappa$ denote the electrical, thermo-electric, electro-thermal and constant voltage thermal conductivity matrix, respectively. 
The thermopower matrix is defined as $S_{ij} = [\sigma^{-1} \alpha]_{ij}$ and the open circuit thermal conductivity is $\kappa_{ij} = [\bar{\kappa} - T \bar{\alpha} \sigma^{-1} \alpha]_{ij}$. The diagonal components of $S$ are the Seebeck coefficients, while the off-diagonal coefficients are the Nernst coefficients. 
Clearly, the different transport coefficients are determined by either the ECA, TCA, GCA, or a combination of two of these (mixed anomalies) \cite{Gooth17}. Explicit analytical expressions of each of the transport coefficient matrix, and their association with different anomalies is discussed in Sec.~S4 of the SM~\cite{Note1}.

We find that the anomaly induced transport coefficients in WSM satisfy Onsager's reciprocity relations, $T\delta \alpha_{ij}(B)  = \delta{\bar \alpha}_{ji}(-B)$ in addition to $\delta {\sigma}_{ij}(B)  =  \delta{\sigma}_{ji}(-B)$ and 
$\delta {\bar \kappa}_{ij}(B)  = \delta{\bar \kappa}_{ji}(-B)$. This is guaranteed by the fact that the ${\cal D}_n^s$ is an even function of ${B}$. The Onsagar's reciprocity relations in WSM have also been shown to be valid in the hydrodynamic transport limit \cite{Lucas16}. However, in contrast to normal metals, the electronic transport coefficients of Eq.~\eqref{currents} violates the Wiedemann-Franz law [$ L ={\kappa}/(T \sigma)={\rm constant}$] as well as the Mott relation [$ M = \alpha/(T \partial_\epsilon \sigma|_\mu) = {\rm constant}$]. 

This summarizes our formulation for exploring magneto-transport in WSM induced by the quantum chiral anomalies. 
This framework can now be combined with {\it ab-initio} based Wannier models, or few orbital based tight-binding models for material specific predictions. Below, we present results for a low energy model of a WSM with a single pair of Weyl nodes.

\label{tilted_WSM}
The low energy Hamiltonian for a Weyl node is 
\be \label{hamiltonian}
\mathcal{H}^s = s \hbar v_F~{\bf \sigma} \cdot {\bf k}~, 
\ee
where $v_F$ is the Fermi velocity and ${\bf \sigma} =\{\sigma_1 ,\sigma_2, \sigma_3\}$ is the set of Pauli spin matrices.
The Berry curvature and the orbital magnetic moment for the conduction band of the Hamiltonian in Eq.~\eqref{hamiltonian}, are ${\bf \Omega}^s = -s {\bf k}/(2 |{\bf k}|^3)$ and 
${\bf m}^s = -s ev_F {\bf k}/(2|{\bf k}|^2)$, respectively (see Sec.~S5 of the SM \cite{Note1} for more details). 
\begin{figure}[t]
\includegraphics[width=.95\linewidth]{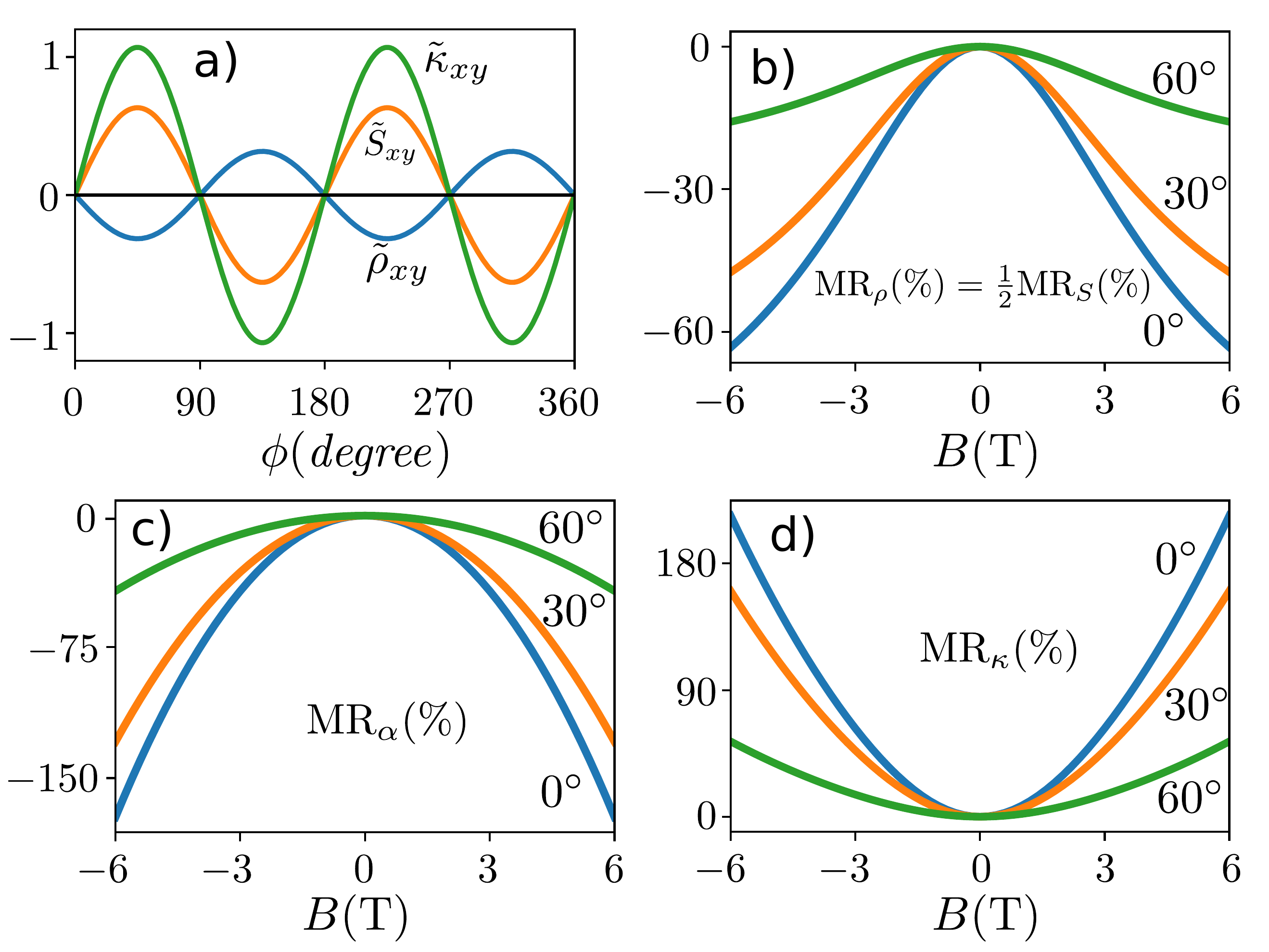}
\caption{Impact of the chiral anomalies induced magneto-transport coefficients in WSM. 
(a) The planar Hall (blue), planar Nernst (orange), and the planar Righi-Leduc (green) effect (normalized by their longitudinal Drude counterpart) are shown as function of the angle $\phi$ between ${\bf B}$ and ${\bf E}$ or $\nabla T$. (b) The saturating negative magneto-resistance (MR$_\rho$) and negetive magneto-Seebeck effect (MR$_S$). (c) The non-saturating negative magneto thermo-electric conductivity (MR$_{\alpha}$) and (d) the positive magneto-thermal conductivity. 
Here we have chosen $T=40$ K, $\mu = 0.05$ eV and $\tau_0 = 10^{-12}$ s, with all the other parameter being identical to those of Fig.~\ref{fig_2}.
\label{fig_3}}
\end{figure}

For this model Hamiltonian, it is straight forward to calculate the quantities defined in Eq.~\eqref{DnVn}. In the limit $\mu > k_B T$, where Sommerfeld expansion is valid, the generalized energy velocities can be calculated to be 
\be \label{Vel_}
\left \{  {\Lambda}_0^s, {\Lambda}_1^s, {\Lambda}_2^s \right \} = -s \dfrac{e }{4\pi^2 \hbar^2} \left \{ {\cal F}_0 , \dfrac{1}{\beta}{\cal F}_1, \dfrac{1}{\beta^2}{\cal F}_2 \right\}. 
\ee
Here, ${\cal F}_i$'s are functions of $ x = \beta \mu$ and we have defined, ${\cal F}_0(x) \equiv 1/(1 + e^{-x}) $, $\mathcal{F}_1(x) \equiv x/(1 + e^x) + \ln[1+e^{-x}]$ and ${\cal F}_2 (x)\equiv \pi^2/3- x \left(\frac{x}{1+e^x}+2\ln\left[1+e^{-x}\right]\right)+ 2 {\rm Li}_2(-e^{-x})$, with {\rm Li} denoting the polylog function (see Fig.~S1 in the SM \cite{Note1} for their temperature dependence).  
Similarly we calculate the generalized energy density of states, 
\be  \label{DOS_}
\left\{{\cal D}_0^s,{\cal D}_1^s,{\cal D}_2^s \right\}  \approx  \dfrac{\mu^2}{2 \pi^2}\dfrac{1}{\hbar^3 v_F^3}\left\{{\cal F}_0,  \dfrac{2}{\beta^2 \mu}{\cal F}_2, \dfrac{1}{\beta^2} {\cal F}_2\right\}.
\ee
Note that in calculating the generalized energy densities, we have neglected the magnetic field correction which is very small. 
In the limiting case of  $\beta \mu \to \infty$, we have $\{{\cal F}_0, {\cal F}_1, {\cal F}_2\} \to \{1, 0,\pi^2/3\}$. In this limit, ${\cal C}_1^s \to 0$ implying that there is no TCA. 

Using Eqs.~\eqref{Vel_}-\eqref{DOS_} in Eq.~\eqref{chrl_qnts}, we evaluate $\delta \mu^s$ and $\delta T^s$ in the $\beta \mu \to \infty$ limit to be 
\be \label{chirl_CP_1}
\begin{pmatrix}
\delta \mu^s \\ \delta T^s/T 
\end{pmatrix}
= s\dfrac{\tau_v \hbar^2 v_F^3}{\mu^2}\frac{1}{2\hbar}
\begin{pmatrix}
\frac{e^2}{2} & -\frac{e k_B}{\beta \mu} \frac{\pi^2}{3} \\
-\frac{e^2}{\mu} & \frac{e}{T}
\end{pmatrix}
\begin{pmatrix}
{\bf B} \cdot {\bf E}\\ {\bf B} \cdot {\bf \nabla}T
\end{pmatrix}~.
\ee
For finite $\beta \mu$, their variation in the $\mu-T$ plane (for $s=1$ node) is shown in Fig.~\ref{fig_2}, and analytical results are presented in Sec.~S4 of the SM \cite{Note1}. The ${\bf B}\cdot{\bf E}$ and ${\bf B}\cdot{\bf \nabla}T$ terms compete with each other to change both $\delta \mu^s$ as well as $\delta T^s$. 


%
%
The chiral anomaly induced transport coefficients ($\sigma,\alpha,\bar{\alpha}$, and $\bar{\kappa}$) can now be obtained from Eq.~\eqref{currents}. In the 
$\beta \mu \to \infty$ limit, these are given by 
\be \label{matrix}
\begin{pmatrix}
\sigma_{ij} &  \alpha_{ij} \\
{\bar \alpha}_{ij} & {\bar \kappa}_{ij}
\end{pmatrix}=  \frac{\tau_v }{2} \frac{e^2v_F^3}{8\pi^2\hbar} \frac{B^2}{\mu^2}
\begin{pmatrix}
e^2 &  \frac{2\pi^2}{3} \frac{ek_B}{\beta\mu}  \\
\frac{2\pi^2}{3}\frac{ek_BT}{\beta\mu} & \frac{\pi^2}{3} k_B^2T
\end{pmatrix} {\cal L}_{ij}(\theta,\phi)~.
\ee
Here, $(\theta,\phi)$ denotes the polar and azimuthal angle of the magnetic field. Note that in Eq.~\eqref{matrix}, $\alpha$, and ${\kappa}$ are $\propto T$ while 
${\bar{\alpha}} \propto T^2$. The angular dependence of all transport coefficients 
is given by the matrix 
\small
{
\be  \label{angular}
{\cal L (\theta,\phi}) \equiv 
\begin{pmatrix}
\sin^2 \theta \cos^2 \phi & \frac{1}{2}\sin^2 \theta  \sin 2\phi & \frac{1}{2}\sin 2\theta \cos \phi \\
\frac{1}{2}\sin^2 \theta \sin2\phi & \sin^2 \theta \sin^2 \phi & \frac{1}{2}\sin 2\theta \sin \phi \\
\frac{1}{2}\sin 2\theta \cos \phi & \frac{1}{2}\sin 2\theta \sin \phi & \cos^2 \theta 
\end{pmatrix}. 
\ee}%
\normalsize
Co-planar ${\bf E}$ and ${\bf B}$, give rise to the planer Hall effect ($\sigma_{xy}$) along with the planar Ettingshausen effect (${\bar S}_{xy} = T S_{xy}$). 
Co-planar ${\nabla T}$ and ${\bf B}$ result in the planar Nernst effect ($S_{xy}$)  and the the planar Righi-Leduc effect ($\kappa_{xy}$). 
We find that the angular dependence of $\kappa_{xy}$ and $S_{xy}$ is identical ($\propto \sin2\phi$), in contrast to $\rho_{xy}$ which is $\propto - \sin 2\phi$ as shown in Fig.~\ref{fig_3}(a).
\begin{figure}[t]
\includegraphics[width=.95\linewidth]{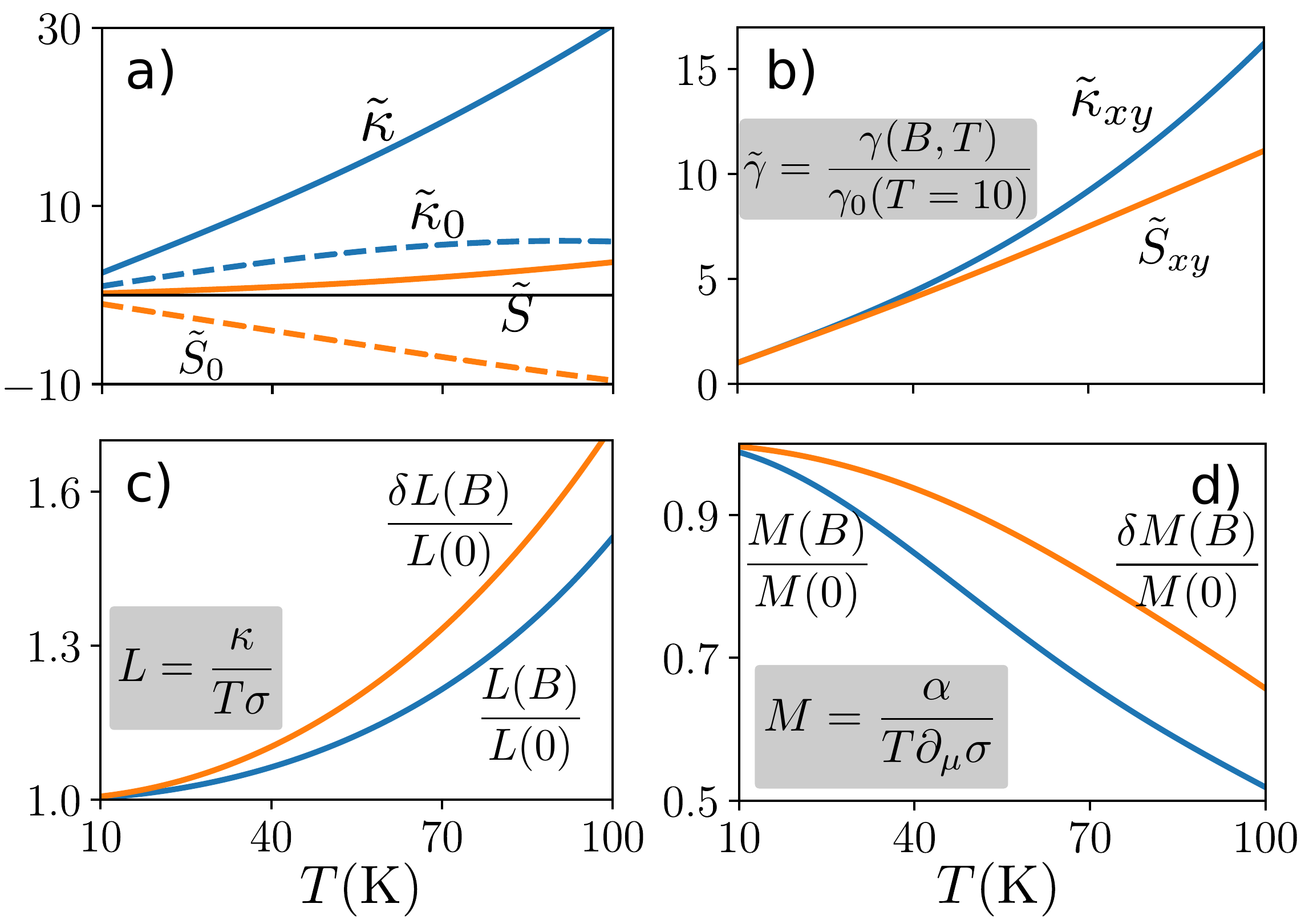}
\caption{(a) The $T$ dependence of Seebeck coefficient and thermal conductivity (scaled by their Drude values at 10K). The Drude $\kappa_0$ (blue dashed line) is positive and GCA enhances it significantly 
(blue solid line). The Drude $S_0$ is negative (orange dashed line) for $\mu > 0$,  but it flips sign in presence of magnetic field (orange solid line). (b) The GCA induced enhancement with temperature is also evident in the planar $\kappa_{xy}$ and $S_{xy}$ (scaled by their values at 10K) when the angle between ${\bf E}$ or $\nabla T$ and ${\bf B}$ is 45$^\circ$. (c) The quantum anomalies induced violation of the Wiedemann-Franz law and (d) the Mott relation. 
Here, all the parameter are identical to those used for Figs.~(\ref{fig_2})-(\ref{fig_3}).
\label{fig_4}}
\end{figure}%

To explore the impact of anomalies on longitudinal magneto-resistance (MR), we define generic MR$_\gamma \equiv \gamma(B)/\gamma(0) - 1$, where $\gamma$ denotes 
the transport coefficients: $\sigma$, $\alpha$, ${\bar \alpha} $, $\kappa$ or $S$. To evaluate the MR, we need to add the Drude components to the respective transport coefficients.  
Evaluating the Drude components for each Weyl node \cite{Das19b}, we have 
%
$\{\sigma_0^s, ~\alpha_0^s,~\bar \kappa_0^s \} = \frac{\mu^2 \tau_0}{6\pi^2\hbar^3 v_F} \Big\{e^2 {\cal F}_0, -2\frac{e k_B}{\beta \mu} {\cal F}_2, \frac{k_B}{\beta}{\cal F}_2 \Big\}.$  
In the $\beta \mu \to \infty$ limit, to all orders in $B$ we obtain the longitudinal MR,
$\left\{{\rm MR}_\rho,~{\rm MR}_S\right\} = -\frac{3 \tau_v \zeta^2}{2 \tau_0+3 \tau_v \zeta^2}~\left\{1, 2\right\}.$
%
Here $\zeta= e \hbar v_F^2 B/(2 \mu^2)$.
Both of these show negative MR which is quadratic for small $B$ and saturates with increasing $B$ values. This behaviour, along with MR$_{\rho}$/MR$_{S} = 1/2$,  persists even for finite $\beta \mu$ values, as shown in Fig.~\ref{fig_3}(b) for an electronically doped system ($\mu >0$). Negative saturating MR$_\rho$ in WSM has been reported in several experiments \cite{Xiong15,Huang15b,Li16a}. Negative MR$_S$ has also been reported in recent experiments on GdPtBi \cite{Hirschberger16} and Cd$_3$As$_2$ \cite{Jia16}.

For $\alpha$ and $\kappa$, in the $\beta \mu \to 0$ limit, we obtain 
$\{{\rm MR}_{\alpha},{\rm MR}_{\bar \kappa}\}  = \frac{3 \tau_v \zeta^2}{2\tau_0}~\{-1,1\}.$
In contrast to MR$_\rho$/MR$_S$ and consistent with calculations of  Ref.~\cite{Spivak16}, both of these show a non-saturating behaviour with MR$_{\alpha}$ being negative, and MR$_{\bar \kappa}$ being positive. This trend persists even for finite $\beta \mu$ as shown in Fig.~\ref{fig_3}(c)-(d). The observation of semi-classical negative MR$_\alpha$ (for $\mu>0$) has also been reported in the Weyl semimetal NbP \cite{Gooth17} as well in GdPtBi \cite{Hirschberger16}. A positive MR$_\kappa$ has been recently reported in Ref.~\cite{Schindler18}. 
We emphasize that the relatively larger MR in $\alpha$ and $\kappa$ as compared to MR$_\rho$, has its origin primarily in the ${\cal C}_{2}^s$ term or the GCA.

The temperature dependence the diagonal and off-diagonal transport coefficients is shown in Fig.~\ref{fig_4}(a)-(b). 
While the Drude Seebeck coefficient is negative for $\mu >0$, the quantum anomalies reverse its sign, making it positive.  
This sign reversal is a distinct signature of quantum anomalies in chiral fluids (See Fig.~S4 and Sec.~S8 of SM~\cite{Note1} for details). The Drude thermal conductivity is positive, and the  chiral anomalies enhance it significantly [by $\sim6x$ at $T = 100$ in Fig.~\ref{fig_4}(a)]. This possibly explains the exciting experimental report of magnetic field induced $300\%$ enhancement of the electronic component of the longitudinal thermal conductivity in the WSM phase of Bi$_{89}$Sb$_{11}$ \cite{Vu19}. We also find a similar enhancement in $S_{xy}$ and $\kappa_{xy}$, as shown in Fig.~\ref{fig_4}(b), and also in $\alpha$ while the relative change in $\rho$ is small 
(see Fig.~S2 in SM~\cite{Note1}). 

The violation of the Wiedemann-Franz law and the Mott relations in the anomaly induced transport coefficients is shown in Fig.~\ref{fig_4}(c)-(d). To highlight that 
this violation goes much beyond the breakdown of the Sommerfeld expansion, which can occur in normal metals as well, we plot the ratio, $L(B,T)/L_0(T)$ and $M(B,T)/M_0(T)$, where $L_0$ and $M_0$ are the corresponding Drude counterpart.  See Sec. S7 of the SM \cite{Note1} for more details. 

To summarize, we have predicted a new quantum anomaly in chiral fluids, the thermal chiral anomaly, which pumps chiral charge from one node to other in presence of ${\bf \nabla}T \cdot {\bf B} \neq 0$. 
We have shown that in addition to planar Hall and planar Nernst effects, the planar Ettinghausen and Righi-Leduc effects also arise from these anomalies. We predict a significant enhancement in the magnetic field induced thermo-electric conductivity, Seebeck effect, Nernst effect and thermal conductivity with increasing temperature. We also demonstrate that the quantum anomaly induced magneto-transport in WSM violates the Wiedemann-Franz law and Mott relation at finite temperatures. Our work provides a robust framework for exploring and interpreting the impact of quantum anomalies in magneto-transport experiments in WSM.


\acknowledgements{AA acknowledges Science Education and Research Board (SERB) and Department of Science and Technology (DST) of government of India for financial support.}
\bibliography{THE_chrl_anmly_v2} 

\end{document}